\documentclass{aastex63}

\usepackage{amsmath}

\shorttitle{}
\shortauthors{}

\begin{document}

\title{Microwave Imaging Spectroscopy Diagnosis of the Slow-rise Precursor of a Major Solar Eruption}

\correspondingauthor{Xin Cheng}\email{xincheng@nju.edu.cn}

\author[0000-0002-4198-2333]{Yuankun Kou}
\affil{School of Astronomy and Space Science, Nanjing University, Nanjing 210023, China\\}
\affil{Key Laboratory of Modern Astronomy and Astrophysics (Nanjing University), Ministry of Education, Nanjing 210093, China\\}
\affil{School of Physics and Astronomy, University of Glasgow, Glasgow G12 8QQ, UK}

\author[0000-0003-2837-7136]{Xin Cheng}
\affil{School of Astronomy and Space Science, Nanjing University, Nanjing 210023, China\\}
\affil{Key Laboratory of Modern Astronomy and Astrophysics (Nanjing University), Ministry of Education, Nanjing 210093, China\\}

\author[0000-0003-2872-2614]{Sijie Yu}
\affil{Center for Solar-Terrestrial Research, New Jersey Institute of Technology, 323 M L King Jr. Blvd., Newark, NJ 07102-1982, USA}

\author[0000-0002-5431-545X]{Yingjie Luo}
\affil{School of Physics and Astronomy, University of Glasgow, Glasgow G12 8QQ, UK}

\author[0000-0002-0660-3350]{Bin Chen}
\affil{Center for Solar-Terrestrial Research, New Jersey Institute of Technology, 323 M L King Jr. Blvd., Newark, NJ 07102-1982, USA}

\author[0000-0002-4978-4972]{Mingde Ding}
\affil{School of Astronomy and Space Science, Nanjing University, Nanjing 210023, China\\}
\affil{Key Laboratory of Modern Astronomy and Astrophysics (Nanjing University), Ministry of Education, Nanjing 210093, China\\}

\begin{abstract}

In this Letter, taking advantage of microwave data from the Expanded Owens Valley Solar Array and extreme-ultraviolet (EUV) data from the Atmospheric Imaging Assembly, we present the first microwave imaging spectroscopy diagnosis for the slow-rise precursor of a major coronal mass ejection (CME) on 2022 March 30. The EUV images reveal that the CME progenitor, appearing as a hot channel above the polarity inversion line, experiences a slow rise and heating before the eruption. The microwave emissions are found to mainly distribute along the hot channel, with high-frequency sources located at the ends of the hot channel and along precursor bright loops underneath the hot channel. The microwave spectroscopic analysis suggests that microwave emissions in the precursor phase are dominated by thermal emission, largely different from the main phase when a significant non-thermal component is present. These results support the scenario that the precursor reconnection, seeming to be moderate compared with the flare reconnection during the main phase, drives the build-up, heating, and slow rise of CME progenitors toward the explosive eruption.

\end{abstract}

\section{Introduction} \label{sec:intro}

Coronal mass ejections (CMEs) are the most energetic phenomena happening in the solar atmosphere, and could have severe impacts on the Earth's ionosphere and modern telecommunication activities. Their predication is therefore of vital importance and requires a thorough comprehension of the initiation and early kinematic evolution. Features observed in extreme-ultraviolet (EUV) and X-ray bands such as sigmoids, filaments, and hot channels etc \citep[see][for a review]{cheng2017_origin} are recognized as progenitors of CMEs. It is now generally acknowledged that these CME progenitors are distinct manifestations of the pre-eruptive magnetic field configuration above the polarity inversion line (PIL) of the photospheric magnetic field in active regions, which are believed to be sheared arcades \citep{sturrock1966_model, antiochos1999_model} or magnetic flux ropes \citep[MFRs;][]{shibata1995_hot-plasma, vourlidas2012_how, cheng2017_origin}. They play an important role in causing the eruption. The key difference between sheared arcades and MFRs is that a central axis exists for the latter case in which all magnetic field lines wrap around it. In the past decades, the properties of pre-eruptive configurations have been investigated extensively with the help of high-resolution EUV imaging \citep{green2007_transient, cheng2013_driver}, H$\alpha$ and EUV/ultraviolet (UV) spectral line analysis \citep{zhou2016_observationsa, awasthi2018_chromospheric, wang2023_spectrala}, and non-linear force-free field (NLFFF) modeling of the magnetic field \citep{yan2001_magnetic, inoue2012_nonlinear, awasthi2018_pre-eruptive}. For individual events, the initiation is often interpreted by mechanisms including tether-cutting reconnection \citep{moore1980_filament} or breakout reconnection \citep{antiochos1999_model, chen2023_observations}, as well as various ideal magnetohydrodynamic (MHD) instabilities \citep{forbes1991_catastrophe, kliem2006_torus}, but which one is universal needs to be further investigated. 

Our recent studies emphasized that a precursor phase of solar flares, corresponding to the transit of the CME progenitors from the quasi-static to impulsive acceleration phase, is able to be used to diagnose the initiation mechanisms of CMEs/flares \citep{cheng2020_initiation,cheng2023_deciphering,xing2024_unveilinga}. During the precursor phase, the CME progenitors rise with a relatively small speed and moderate acceleration compared with that in the main phase. With hard X-ray (HXR) imaging spectroscopy and EUV observations, \citet{cheng2023_deciphering} suggested that a precursor magnetic reconnection is responsible for the heating and early slow rise of the pre-eruptive MFR. After the MFR reaches a critical height, the torus instability occurs and then joins with the reconnection to continue to drive the slow rise for a while until the start of the flare reconnection, as subsequently confirmed by the 3D thermal-MHD simulation of \citet{xing2024_unveilinga}. Besides, the precursor reconnection also gives rise to energetic electrons, but which seems to be unimportant given the HXR spectra are dominated by the thermal component \citep{cheng2023_deciphering}. 

Recently, microwave imaging spectroscopy observations have provided new tools to investigate eruptive MFRs and their dynamics. For instance, using imaging spectroscopy data obtained by the Expanded Owens Valley Solar Array (EOVSA; \citealt{gary2018_microwave}) in 1--18 GHz, \citet{chen2020_microwave} observed microwave emissions near the two anchor ends of an MFR during its early-impulsive eruption phase, which exhibit temporal and spatial behavior similar to the primary reconnection source. It documents the possibility of microwave imaging spectroscopy data that can be potentially used to diagnose the physical processes during the precursor phase. Moreover, microwave imaging spectroscopy observations from EOVSA also show powerful ability to diagnose the coronal magnetic field and electron energy distributions during flares \citep[e.g.,][]{yu2020_magnetic, chen2020_measurement, wei2021_coronal, fleishman2022_solar}, and thus also hold the potential of quantifying these properties for the precursor phase. Using total-power microwave data from the Nobeyama Radio Polarimeters (NoRP) and Radio Solar Telescope Network (RSTN), \citet{altyntsev2012_thermal} found non-thermal emissions and estimated the magnetic field strength at the early stage of several flares. Similarly, with the total-power data provided by EOVSA (before it was commissioned to perform imaging spectroscopy), \citet{wang2017_highresolution} and \citet{liu2022_multiinstrument} derived the temporal evolution of the averaged magnetic field strength during the two consecutive precursor stages of a major flare. Nevertheless, it should be pointed out that these studies are based on microwave data without spatial resolution and thus are difficult to definitively determine the nature of pre-eruptive activities.

In this Letter, we present the first microwave imaging spectroscopy diagnosis for the slow rise of pre-eruptive structure of a major eruption on 2022 March 30. It is found that the pre-eruptive structure, appearing as a hot channel, experienced a slow expansion and rise before the eruption. Interestingly, the microwave emissions distributed along the hot channel, with high-frequency components concentrating near its two ends and the precursor loops. Utilizing spatially resolved microwave spectra, the energy distributions of associated electrons within the precursor structures, as well as their magnetic field strength, are estimated. The results support the scenario that the pre-eruptive hot channel is experiencing a moderate reconnection that is largely different from the flare reconnection in the main phase. In Section \ref{sec:obs}, we briefly introduce instruments and the data reduction including the gyrosynchrotron-model fitting of microwave spectra. In Section \ref{sec:results}, we present observational results and analyses, which are followed by a summary and discussion in Section \ref{sec:sum}.

\section{Instruments and Data Reduction} \label{sec:obs} 

The Atmospheric Imaging Assembly \citep[AIA; ][]{lemen2012_atmospheric} on board Solar Dynamics Observatory (SDO) provides high-resolution EUV (UV) images (0.6\,arcsec/pixel) with high cadence of 12\,s (24\,s) at 9 passbands. Because of being sensitive to high-temperature plasma, the 94 {\AA} and 131 {\AA} images are used to detect pre-eruptive hot channels and precursor hot loops. The 1600 {\AA} images are utilized to determine their footpoints, with a combination of the photospheric magnetograms from Helioseismic and Magnetic Imager \citep[HMI; ][]{scherrer2012_helioseismic}. 

Based on AIA images at six EUV bands (i.e., 94, 131, 171, 193, 211, and 335 {\AA}), we conduct the differential emission measure (DEM) analysis to estimate the density and temperature of pre-eruptive structures. We adopt the procedures proposed by \citet{cheung2015_thermal} and \citet{su2018_determination}, where the images are re-binned to 1.2 arcsec/pixel and averaged over every two instances to increase the signal-to-noise ratio. 

EOVSA images the Sun with a temporal resolution of 1 s in 1--18 GHz with 451 spectral channels. During the observation period, the spectral channels were combined into 50 spectral windows (SPWs) for the sake of increasing the dynamic range. In this work, we use SPW 3 to 50 (2.8 to 18 GHz) with a bandwidth of 325 MHz for each SPW. The synthesized beam size at the observation period is $74^{\prime \prime}.4/\nu\mathrm{(GHz)} \times 52^{\prime \prime}.9/\nu\mathrm{(GHz)}$. When reconstructing microwave images, we set circular restoring beam sizes at $88^{\prime \prime}/\nu\mathrm{(GHz)}$ and use the {\texttt{CLEAN}} algorithm \citep{hogbom1974_aperture} to reduce the effects of sidelobes. All routines can be found in the {\texttt{CASA}} \citep{thecasateam2022_casa} and {\texttt{SunCASA}} (\url{https://github.com/suncasa/suncasa-src}) packages. 

Taking advantage of microwave images at multiple frequencies, we derive spatially resolved brightness temperature spectra $T_{\mathrm{b}} (\nu)$ for the particular structures. To derive their plasma parameters and magnetic field strength, we perform a gyrosynchrotron-model fitting. The {\texttt{Fast Gyrosynchrotron Codes}} developed by \citet{fleishman2010_fast} and \citet{kuznetsov2021_ultimate} are used to calculate the microwave spectra from given parameters. The non-linear least chi-square method involved in the {\texttt{scipy.optimize}} package is adopted to find the best fitting of calculated spectra to observed ones. The {\texttt{emcee}} package \citep{foreman-mackey2013_emcee} is utilized to conduct Markov Chain Monte Carlo (MCMC) simulation to evaluate the fitting results. In our fitting process, we assume the electron energy distribution either thermal Maxwellian contributing on both free-free and thermal gyrosynchrotron radiation or its combination with a non-thermal power-law form ($f(\epsilon) \sim {\epsilon}^{- \delta}$, where $\epsilon$ is the electron energy and $f$ is the electron density distribution function) contributing on non-thermal gyrosynchrotron radiation. The non-thermal distribution is set with a low-energy cut-off 10\,keV and a high-energy cut-off 10\,MeV. The column depth for microwave radiation is set as the approximate apparent width of the structures, 18\,arcsec, as seen from the EUV image (e.g., Figure \ref{fig:overview}a). There are four free parameters remaining for the model fitting using a pure-thermal model: magnetic field strength $B$, plasma temperature $T$, density of thermal electrons ${n_{\mathrm{th}}}$, and the angle $\theta$ between the magnetic field direction and the line of sight (LOS). And for the model with non-thermal electrons, density of non-thermal electrons ${n_{\mathrm{nth}}^{>10\,{\mathrm{keV}}}}$ and the power-law index of non-thermal electrons $\delta$ are two additional free parameters. The MCMC fitting process gives the posterior probability distributions (PPDs) of the parameters in the form of a ``corner plot'' as shown in Appendix Figures, where the best fitting results are labeled by blue solid lines passing through each panel, in rows and in columns, with the red lines representing the fitting results from the non-linear least chi-square method.

\section{Results} \label{sec:results}

\subsection{Event Overview}

\begin{figure*}
\gridline{\fig{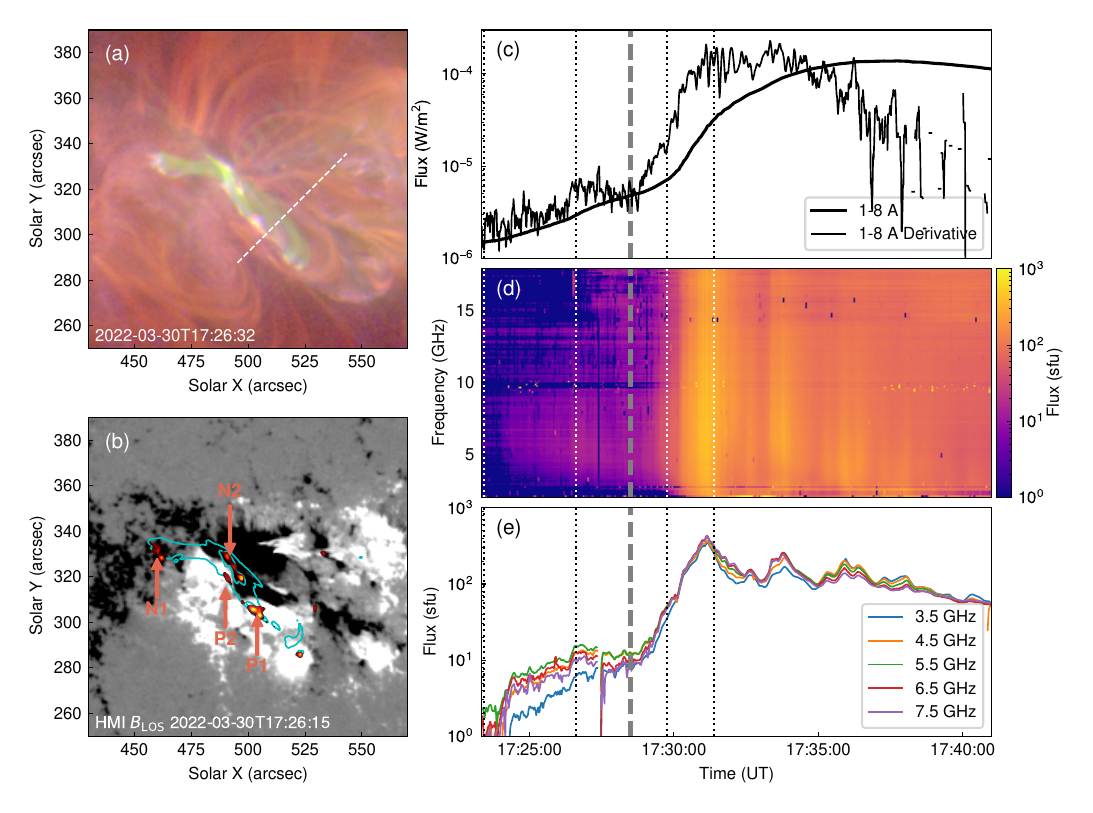}{0.8\textwidth}{}}
\caption{(a) Composite of the AIA 131 {\AA} (green), 171 {\AA} (red), and 304 {\AA} (blue) image showing the pre-eruptive hot channel. The white dashed line labels the slit used for creating the height-time plot in Figure \ref{fig:slit}. (b) The HMI $B_{\mathrm{LOS}}$ magnetogram covered with the 131 {\AA} contours (in cyan, at the levels of 70\% and 90\%, 17:26:32 UT) and 1600 {\AA} contours (in dark red, greater than 20\% of the maximum value within the field of view shown on the panel at 17:26:40 UT). Positive (P1 and P2) and negative footpoints (N1 and N2) are labeled in orange. (c) GOES SXR 1--8\,{\AA} curve and its derivative curve (5\,s smoothed). (d) EOVSA total-power dynamic spectrum during the flare. (e) Microwave flux curves at five frequencies (5\,s smoothed), extracted from the above total-power dynamic spectrum. The vertical dashed lines in panels (c)-(e) label the approximate time (around 17:28:30 UT) when the precursor phase transits to the main phase. The vertical dotted lines mark the four time instances as in Figure \ref{fig:images}.}\label{fig:overview}
\end{figure*}

The eruption of interest on 2022 March 30 originated in the active region 12975. As shown by the composite of three AIA images (131, 171, and 304 {\AA}) at about 17:26:32 UT, a bright elongated structure appeared before the eruption (green component, also in Figure \ref{fig:overview}a) and was wrapped by surrounding loops as seen in 171 {\AA} (red) and 304 {\AA} (blue) bands. The HMI LOS magnetogram indicates that the pre-eruptive bright structure lies above and approximately along the main PIL.

In terms of the soft X-ray (SXR) flux detected by the Geostationary Operational Environmental Satellite (GOES), the X1.4 class flare can be divided into three phases: the precursor phase, the main phase and the decay phase, the first two of which corresponded to the slow rise and fast acceleration of the CME eruption, respectively. The flare precursor started at 17:21 UT and exhibited an enhanced emission afterwards (Figure \ref{fig:overview}c), similar to many previous observations \citep[e.g.,][]{bumba1959_chromospheric,asai2006_preflare, hudson2020_hot}. Around 17:28:30 UT, the main phase of the flare started with the SXR emission increasing rapidly, reaching the peak at 17:37 UT, with a rise time of about 8.5 minutes. Interestingly, the background-subtracted total-power microwave dynamic spectrum (DS) provided by EOVSA also observed a group of weak microwave bursts during the precursor phase (Figure \ref{fig:overview}d), which even presented multiple peaks in microwave flux curves as shown in Figure \ref{fig:overview}e, even though they were about two orders of magnitude weaker than the peak during the main phase.

\subsection{Kinematic Evolution of Pre-eruptive Hot Channel}
\begin{figure*}[ht]
\gridline{\fig{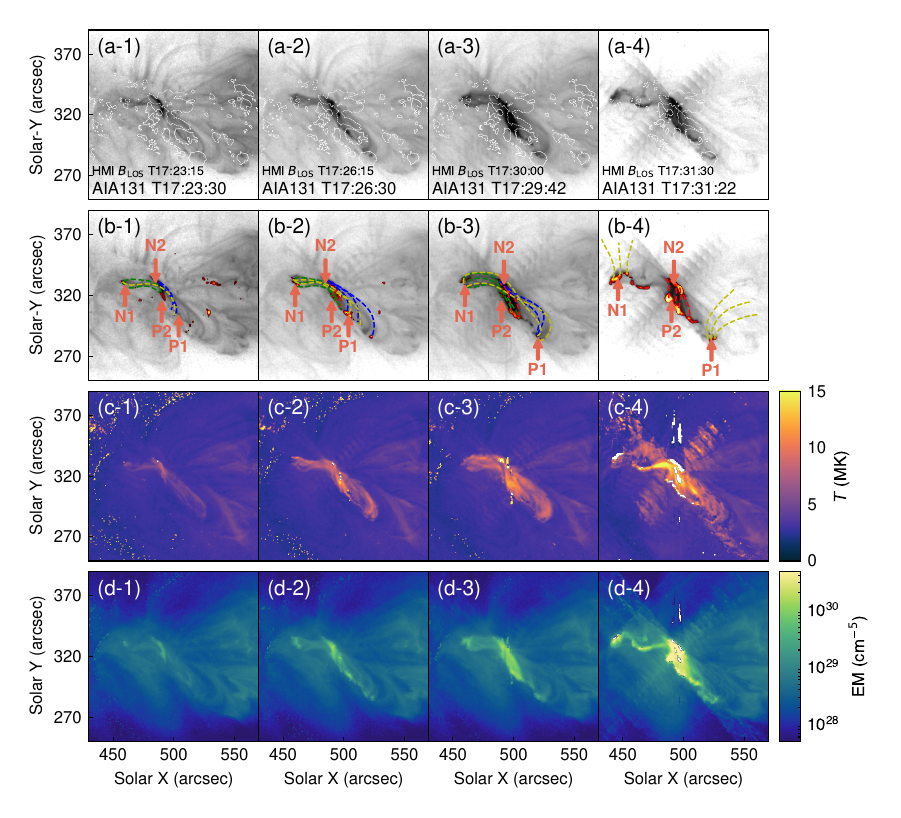}{0.9\textwidth}{}}
\caption{(a) AIA 131 {\AA} images showing the evolution of the hot channel. The color is reversed to highlight the hot channel. The solid and dashed white contours shows the LOS magnetic field of +500\,G and -500\,G, respectively. (b) Same as panels (a) but covered by the 1600 {\AA} brightenings (in dark red, greater than 20\% of the maximum value) to highlight the footpoints of the hot channel and precursor loops, which are delineated by the dashed lines in different colors. The green (blue) lines are sheared magnetic field lines connecting N1 and P2 (N2 and P1), the red lines are the reconnected precursor/flare loops, and the yellow lines are the reconnected hot channel or the erupting MFR. (c) and (d) DEM-averaged temperature and total EM maps. }\label{fig:euv}
\end{figure*}

The AIA 131\,{\AA} images clearly show the early kinematic evolution of the pre-eruptive hot channel. During the precursor phase, the hot channel slowly rose and expanded with the morphology almost unchanged (Figure \ref{fig:euv}a). The average temperature maps in Figure \ref{fig:euv}c show that the emissions above 5\,MK are mainly from the pre-eruptive hot channel, justifying the argument that the hot channel could be an MFR under formation through magnetic reconnection \citep[e.g.,][]{cheng2023_deciphering}. The reconnection most likely took place between two groups of sheared loops as delineated by the dashed lines in green and blue in Figure \ref{fig:euv}b, building up the pre-eruptive hot channel (yellow) connecting footpoints P1 and N1 and low-lying precursor loops (red) connecting footpoints P2 and N2. At the same time, the reconnection heated the plasma, resulting in the radiation of the hot channel and precursor loops at the high-temperature bands. 

\begin{figure}
\gridline{\fig{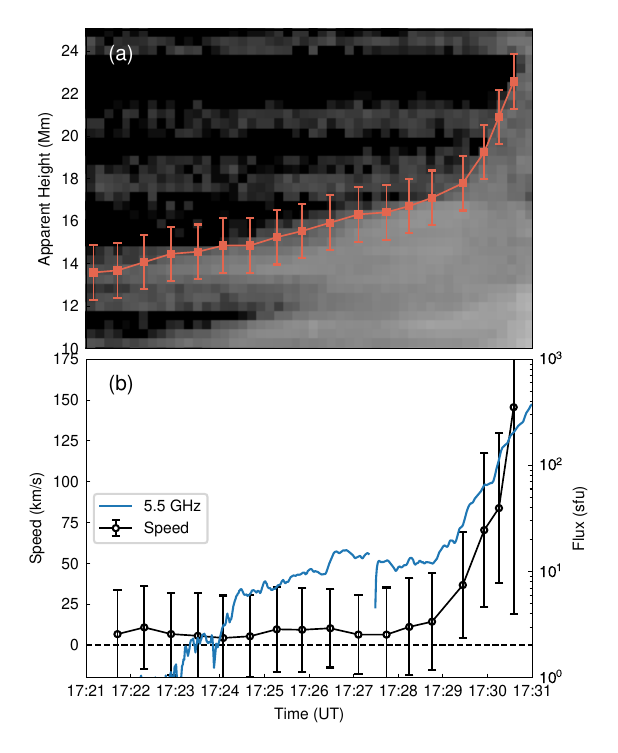}{0.45\textwidth}{}}
\caption{(a) Height-time plot of the AIA 131 {\AA} images. Orange points mark the leading front of the rising hot channel. The errors are set as five pixels (2.18\,Mm). (b) Speed-time profile calculated from the height-time profile in panel (a). A microwave flux curve at 5.5\,GHz is overplotted as the thin blue curve.}\label{fig:slit}
\end{figure}

Figure \ref{fig:slit} shows the temporal evolution of the height and velocity of the hot channel. It is obvious that the velocity almost kept a constant of about 10\,km\,s$^{-1}$ during the entire precursor phase. The slow rise did not stop until the onset of the impulsive acceleration of the hot channel. In the field of view of AIA images, its velocity quickly increased from 10\,km\,s$^{-1}$ to 150\,km s$^{-1}$ in the following two minutes. Note that these velocities are significantly underestimated because of the projection effect.

After the eruption, the hot channel caused a fast CME and a group of flare loops with their footpoints forming two hooked ribbons, as shown in Figure \ref{fig:euv} (the fourth column).

\subsection{Locating Microwave Sources}

\begin{figure*}[ht]
\gridline{\fig{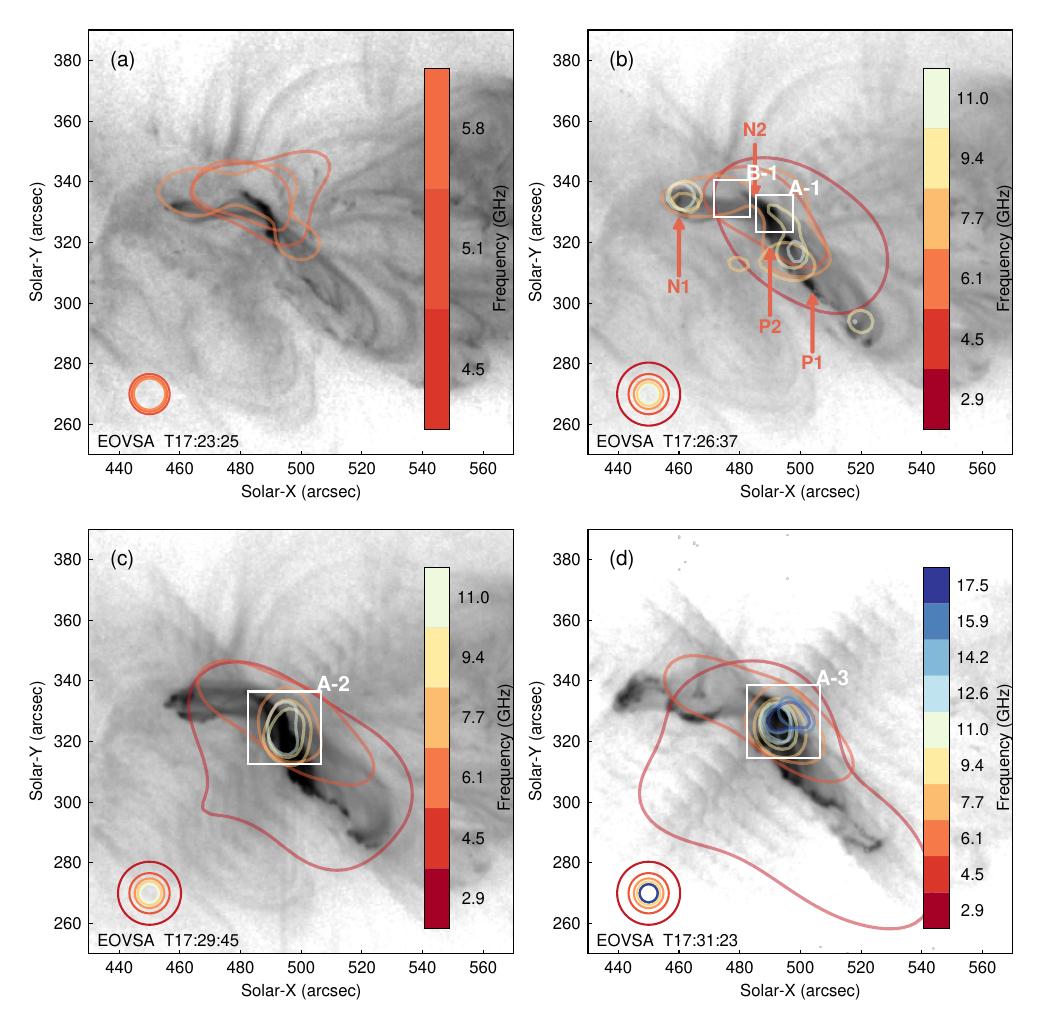}{0.8\textwidth}{}}
\caption{Microwave sources covering AIA 131 {\AA} images. The contours of the microwave emissions are at the levels of 50\% of the maximum brightness temperature at each SPW and at each time. The restoring beam sizes are labeled in the lower left corner of each panel. White boxes represent precursor loops (A-1), hot channel leg (B-1), and flare loops (A-2 and A-3), respectively. The footpoints of the hot channel and precursor loops are annotated with the same symbols as in Figure \ref{fig:euv}b-2. For better presenting microwave emissions from the precursor and flare structures, the contribution from the background active region, taken between 17:22 UT and 17:23 UT, is subtracted when reconstructing images. The average time for obtaining all microwave images is taken as 4\,s.} \label{fig:images}
\end{figure*}

The microwave emissions during the precursor phase are found to mainly originate from the pre-eruptive hot channel and precursor loops. Figure \ref{fig:images} shows the contours of microwave emissions that are overlaid with the AIA 131 {\AA} images. At the early instance (Figure \ref{fig:images}a), the microwave emissions at frequencies between 4.5 and 5.8\,GHz are shown to be along the pre-eruptive hot channel, approximately from the eastern footpoint (N1) to the central bright loops. While for the higher frequency bands, the emissions were too faint to obtain good-quality images. At the precursor peak time (Figure \ref{fig:images}b), however, the microwave emissions became stronger so that the sources above 6.1\,GHz and below 11\,GHz are revealed. The high-frequency emissions were primarily concentrated above the footpoint regions of the hot channel and precursor loops during the precursor phase, but shifted to the central brightening region as entering the main phase (Figure \ref{fig:images}c,d). At the microwave peak time during the main phase, the microwave emissions at frequencies higher than 11\,GHz also appeared (Figure \ref{fig:images}d) and presented a much higher brightness temperature (Figure \ref{fig:spectra}d). It is worth noting that the hot channel had ejected at this instance, meaning that the extensive microwave emissions at lower frequencies were more likely from the post-flare loops, different from those mostly originating from the hot channel during the precursor phase. This could be due to the accelerated electrons producing the microwave emissions during the precursor phase being mainly trapped in the pre-eruptive hot channel. While most of the accelerated electrons in the main phase were trapped in the post-reconnection loops.

\subsection{Properties of Spatially Resolved Microwave Emissions}

To investigate the properties of microwave emissions of the pre-eruptive hot channel, we obtain spatially resolved microwave spectra and perform gyrosynchrotron-model fittings at the hot channel leg region (Box B-1 in Figure \ref{fig:images}b) and associated precursor loops (Box A-1) during the precursor phase, respectively. For comparison, the spectra for the flare loops, likely evolving from the precursor loops, at the beginning (Box A-2) and peak time (Box A-3) of the main phase are also analyzed. The least-square method and MCMC fitting results are presented in the Appendix, and the parameters obtained from MCMC fittings are summarized in Table \ref{tab:fittings}. Note that, due to the optically-thick nature of microwave emissions at the frequencies below the spectrum peak, all spectral analyses are based on the original microwave data without background-subtraction. For reference, the microwave spectrum for the background active region, taken between 17:22 UT and 17:23 UT, as well as the spectral fitting results is shown in Appendix Figure \ref{fig:appendix1}.

Figures \ref{fig:spectra}a and \ref{fig:spectra}b present the spectra as well as their best fitting results for the precursor phase. The spectra at both regions are fitted well with a pure-thermal model involving both thermal gyrosynchrotron and thermal free-free contributions. No apparent distinctions are found between the precursor loops and hot channel leg. The average magnetic field strength of both is around 500--600\,G. The temperature is around 3.8\,MK, and the density of thermal electrons is at the level of $\sim 10^{10}\,{\mathrm{cm}}^{-3}$. The temperature values are approximately at the same level as what are derived by the DEM analyses above (Figure \ref{fig:euv}c and Table \ref{tab:fittings}). 

The similarity of the microwave spectra of the hot channel and precursor loops might be caused by the limitation of the microwave imaging ability at low frequencies. On the one hand, the larger beam sizes for the low-frequency emissions make them less resolved, resulting in the brightness temperature being underestimated (known as “beam dilution”). To reduce the impact from such an uncertainty, when extracting the spatially resolved spectra and performing the fitting, a set of frequency-dependent factors are manually multiplied on the errors that are proportional to the reciprocal of the frequency. 
On the other hand, due to the face-on view, the A-1 source could represent a mixture of the precursor loops and the overlying hot channel. Although the high-frequency emissions within A-1 (e.g., at 9.4 GHz) are shown to be dominated by the precursor loops (Figure \ref{fig:images}b), the low-frequency ones likely include contributions from both structures, thus contaminating the spectrum. As a result, the fitting parameters for A-1 are largely impacted by the overlying hot channel emissions.

By contrast, once entering the main phase, non-thermal electrons appeared (Figure \ref{fig:spectra}c); and presented a power-law energy distribution with the spectral index ($\delta$) becoming smaller as the time elapses, even reaching 2 at the peak time (Figure \ref{fig:spectra}d). Meanwhile, the temperature of thermal electrons also increased largely, much higher than that during the precursor phase (about 14\,MK at the beginning of the main phase; Figure \ref{fig:spectra}c), and increasing to around 60\,MK at the peak (Figure \ref{fig:spectra}d). This shows that a large amount of electrons were accelerated and heated in the main phase. Note that, the microwave emissions during this phase were mainly from post-flare loops; while in the precursor phase, part of them also resulted from the hot channel. No matter what, the different energy distributions of energetic electrons during the precursor and main phase support the two distinct reconnection processes (probably moderate precursor and fast flare reconnection) being in operation in the two different phases, respectively, as proposed by \citet{cheng2023_deciphering}.

\begin{figure*}[ht]
\gridline{\fig{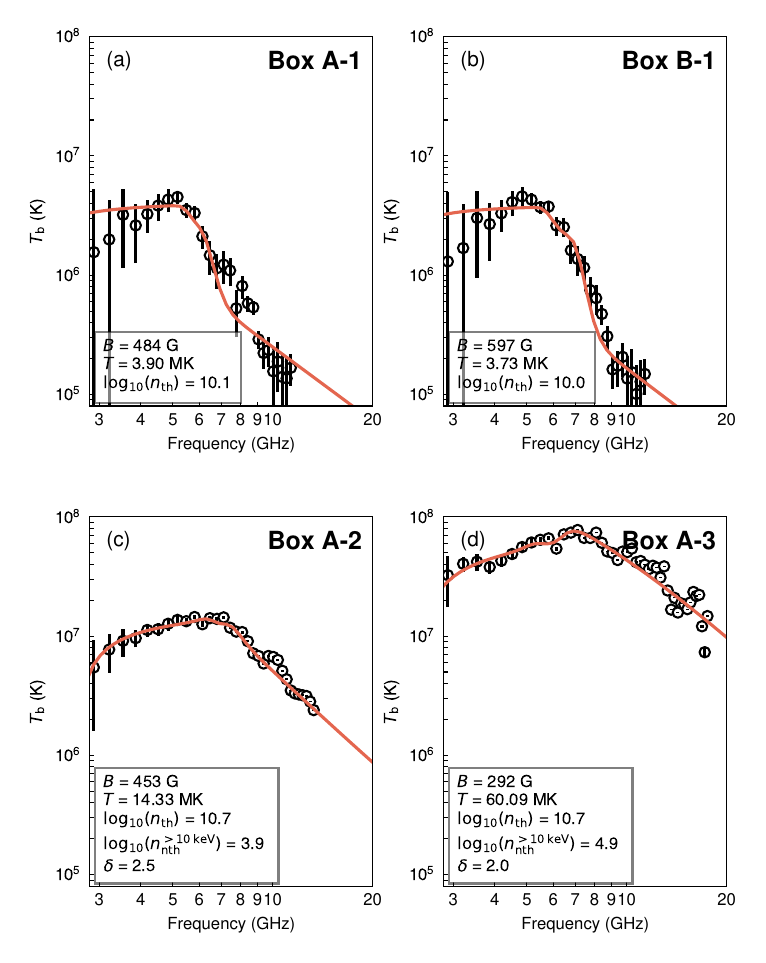}{0.7\textwidth}{}}
\caption{Spatially resolved $T_{\mathrm{b}}$ spectra with the least-square fitting results. The black data points and errors are from the observations, and the red lines are the best fittings with the least chi-square residuals. The main fitting parameters are indicated at each panel. (a) Results for the precursor loops (Box A-1). (b) Results for the hot channel leg (Box B-1). (c) and (d) For the post-flare loops at the beginning (Box A-2) and peak time (Box A-3) of the main phase.
The brightness temperature of each spectrum is obtained by averaging nine brightest pixels in each box to increase the signal-to-noise ratio. The errors are defined as the root-mean-square brightness temperature of a quiet region far from the flare region, multiplied with manually-set frequency-dependent factors. The factors are set for reducing the effect from the different beam sizes at different frequencies.}\label{fig:spectra}
\end{figure*}

\section{Summary and Discussion} \label{sec:sum}

Here, we perform multi-wavelength analyses for the precursor slow rise of a pre-eruptive MFR. The EUV images reveal that the pre-eruptive MFR appeared as a hot channel structure above the PIL and experienced a slow expansion and rise, which afterwards gave rise to a fast CME and an X1.2-class flare. We report the first microwave emissions of the hot channel prior to the eruption and find that they mostly appeared at frequencies below around 6\,GHz and were distributed along the hot channel. The high-frequency emissions presented at the ends of the hot channel, as well as the low-lying precursor loops. 

Moreover, taking advantage of spatially resolved spectral model fitting, we diagnose the magnetic field strength and the energy distribution of energetic electrons associated with pre-eruptive structures. However, the fitting parameters are found to be similar between the A-1 and B-1 sources. This might be due to the face-on view of this event and/or the beam dilution effect. That is to say that the low-frequency emissions within A-1 are contributed from both the precursor loops and the overlying hot channel. The precursor loops and the hot channel are thus not distinguishable for the current event from the perspective of the microwave spectral analysis.

Nevertheless, our observations disclose that the microwave emissions during the precursor phase are mainly produced by thermal electrons, indicating that the plasma heating is dominating the magnetic energy release in this early stage, consistent with the previous observations \citep[e.g.,][]{battaglia2009_observations, cheng2023_deciphering}. This thermal-processes-dominated phase stops once the MFR starts to erupt outward rapidly. Apart from the significant plasma heating, a number of electrons are accelerated to a high energy. We attribute such a transition to the variation of magnetic reconnection schemes. During the precursor phase, two groups of highly-sheared magnetic field lines form a hyperbolic flux tube (HFT) configuration, in which the precursor reconnection takes place. Because of the strong guide field within the HFT, the reconnection could be moderate \citep[e.g.,][]{leake2020_onset, dahlin2022_variability}, leading to the acceleration of electrons being not that efficient. However, the precursor reconnection is able to gradually build up the pre-eruptive MFR, whose two footpoints correspond to the two farther brightening sites, and to cause its slow rise. Once the MFR reached a height where torus instability is triggered, the HFT configuration is quickly stretched so as to form a long current sheet. The flare reconnection then commences, on the one hand, to quickly accelerate the eruption, and on the other hand, to accelerate electrons and heat the plasma efficiently, giving rise to a flat energy distribution of non-thermal electrons.

Moreover, we also estimate the magnetic field strength of the pre-eruptive structure, which was around 500--600\,G., close to some previous estimations by a similar methodology \citep{wang2017_highresolution, wei2021_coronal} or non-linear force-free field extrapolations \citep{korsos2024_first}.

\acknowledgments 
X.C., Y.K., and M.D. are supported by the National Key R\&D Program of China under grants 2021YFA1600504 and the Fundamental Research Funds for the Central Universities under grants 2025300318. The Expanded Owens Valley Solar Array (EOVSA) was designed, built, and is now operated by the New Jersey Institute of Technology (NJIT) as a community facility. EOVSA operations are supported by NSF grant AGS-2436999 to NJIT.

\providecommand{\noopsort}[1]{}

\appendix

\renewcommand{\thetable}{A\arabic{table}}

\setcounter{table}{0}
\begin{table*}[ht]
\caption{Fitting parameters from MCMC simulations.}\label{tab:fittings}
    \centering
    \begin{tabular}{c|ccccccc}
    \tableline \tableline
         Region  &$B$ (G)  &$T$ (MK) &$T_{\mathrm{DEM}}$ (MK)  &$\log_{10}{n_{\text{th}}}\ ({\text{cm}}^{-3})$  &$\log_{10}{n_{\text{nth}}^{>10 \text{keV}}}\ ({\text{cm}}^{-3})$  &$\delta$  &$\theta$\\
         \hline
         Background AR &$611_{-43}^{+60}$  &$1.65_{-0.12}^{+0.13}$  & -  &$10.19_{-0.01}^{+0.01}$  &-  &-  &$60_{-19}^{+19}$\\
         \hline
         A-1  &$508_{-44}^{+75}$  &$3.86_{-0.38}^{+0.41}$  & 4.32 &$10.13_{-0.02}^{+0.02}$  &-  &-  &$52_{-11}^{+13}$\\
         \hline
         B-1  &$591_{-40}^{+37}$  &$3.74_{-0.28}^{+0.30}$  & 3.26 &$10.03_{-0.03}^{+0.03}$  &-  &-  &$46_{-6}^{+8}$\\
         \hline
         A-2  &$417_{-7}^{+8}$  &$13.86_{-0.90}^{+0.87}$  & - &$10.66_{-0.11}^{+0.07}$  &$4.21_{-0.50}^{+0.44}$  &$2.64_{-0.31}^{+0.27}$  &$51_{-4}^{+4}$\\
         \hline
         A-3  &$293_{-7}^{+6}$  &$59.71_{-2.72}^{+3.25}$  & - &$10.70_{-0.06}^{+0.06}$  &$4.80_{-0.36}^{+0.39}$  &$1.99_{-0.17}^{+0.18}$  &$43_{-2}^{+2}$\\
         \hline     
    \end{tabular}
\end{table*}

\renewcommand{\thefigure}{A\arabic{figure}}

\setcounter{figure}{0}

\begin{figure*}[ht]
\gridline{\fig{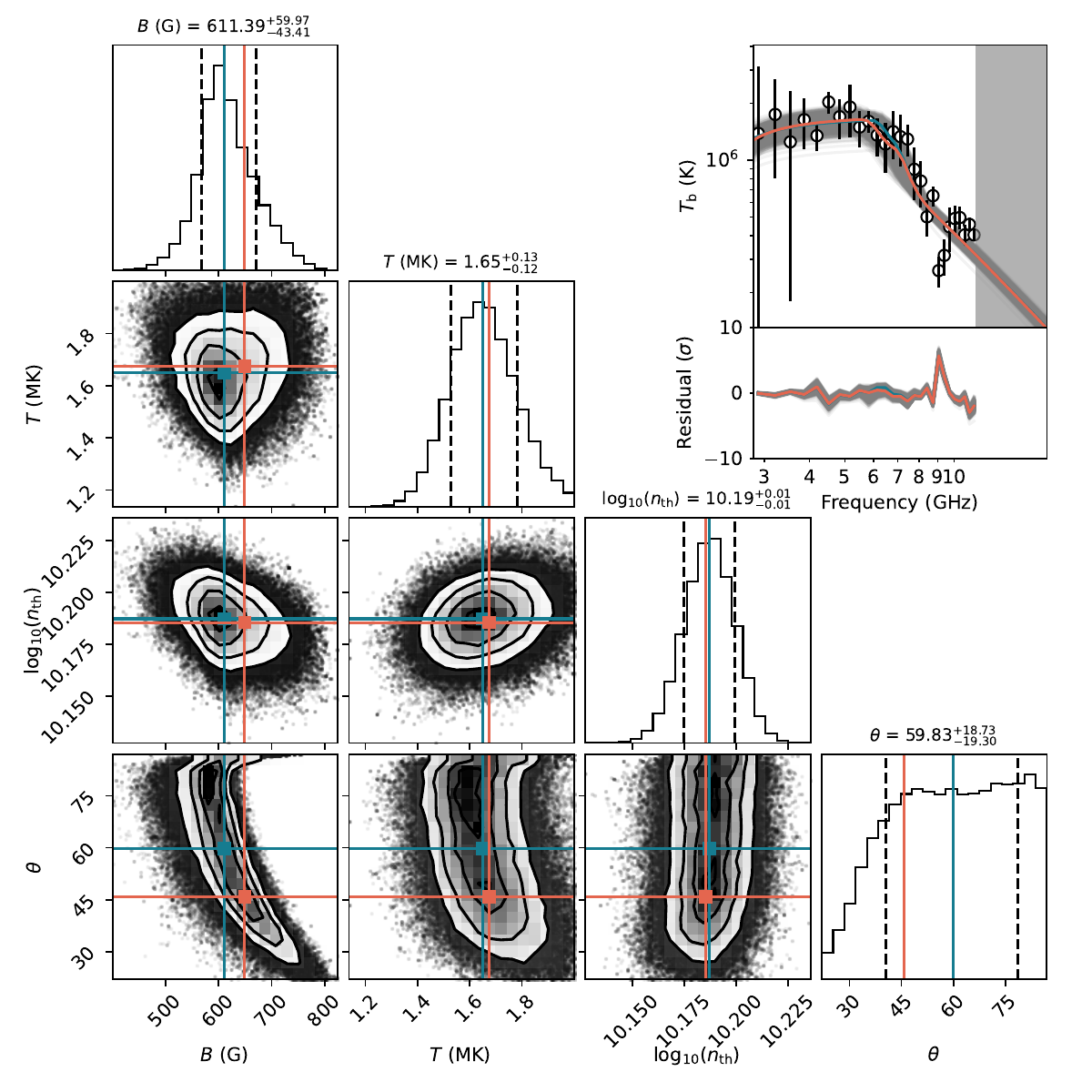}{1\textwidth}{}}
\caption{Posterior probability distributions (PPDs) of the MCMC analysis for the microwave spectrum of the background active region that is derived through averaging the data between 17:22 UT and 17:23 UT. The red lines represent the fitting results by non-linear least chi-square method. While the blue lines are results from the MCMC simulation. The errors are the 1-{$\sigma$} confidence intervals of the PPDs. The gray curves in the upper right panels are spectra and residuals associated with results from MCMC samplings shown in the corner plot.}\label{fig:appendix1}
\end{figure*}

\begin{figure*}[ht]
\gridline{\fig{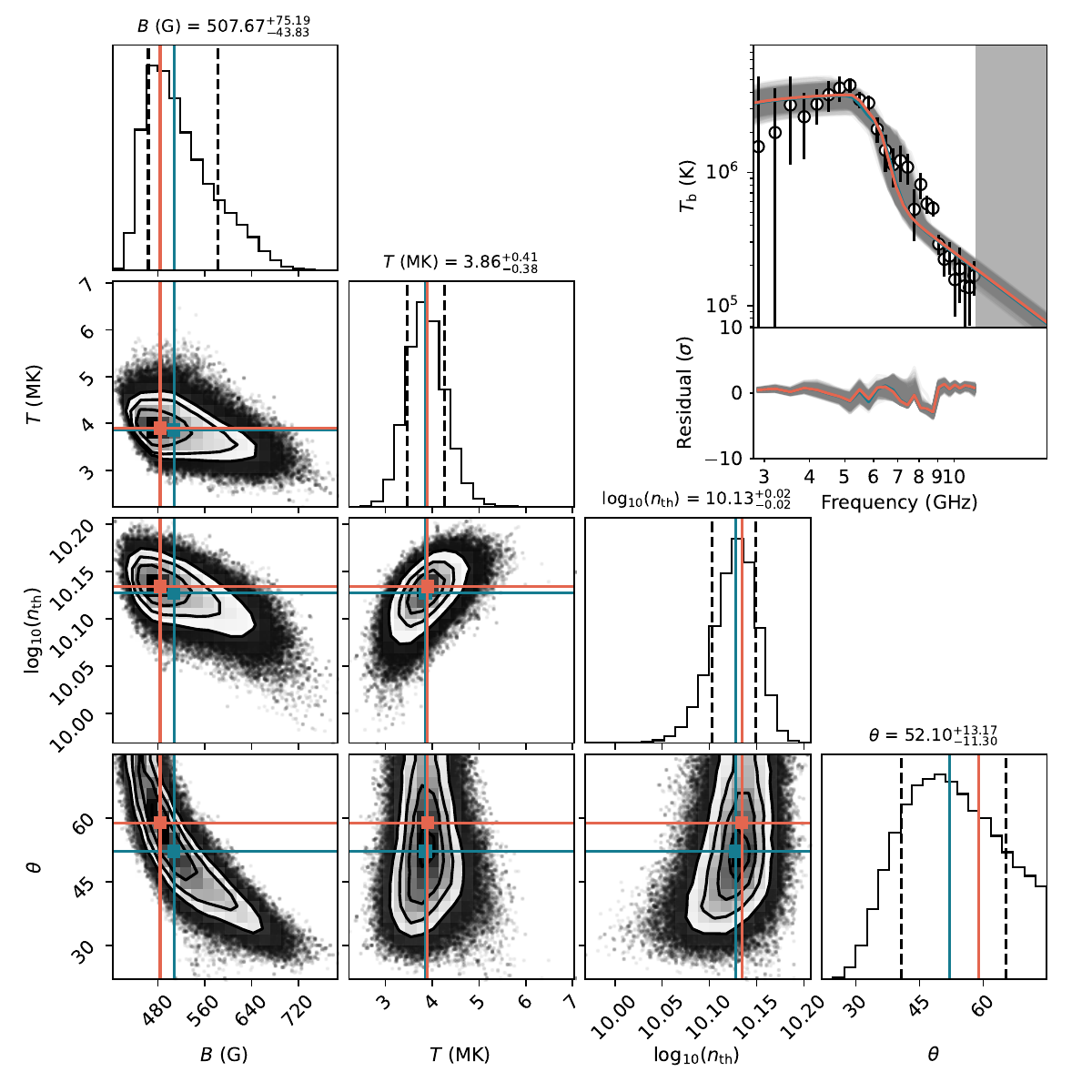}{1\textwidth}{}}
\caption{PPDs of the MCMC analysis for the spectrum from Region A-1 (17:26:36 UT). The red lines represent the fitting results by non-linear least chi-square method. While the blue lines are results from the MCMC simulation. The errors are the 1-{$\sigma$} confidence intervals of the PPDs. The gray curves in the upper right panels are spectra and residuals associated with results from MCMC samplings shown in the corner plot.}
\end{figure*}

\begin{figure*}[ht]
\gridline{\fig{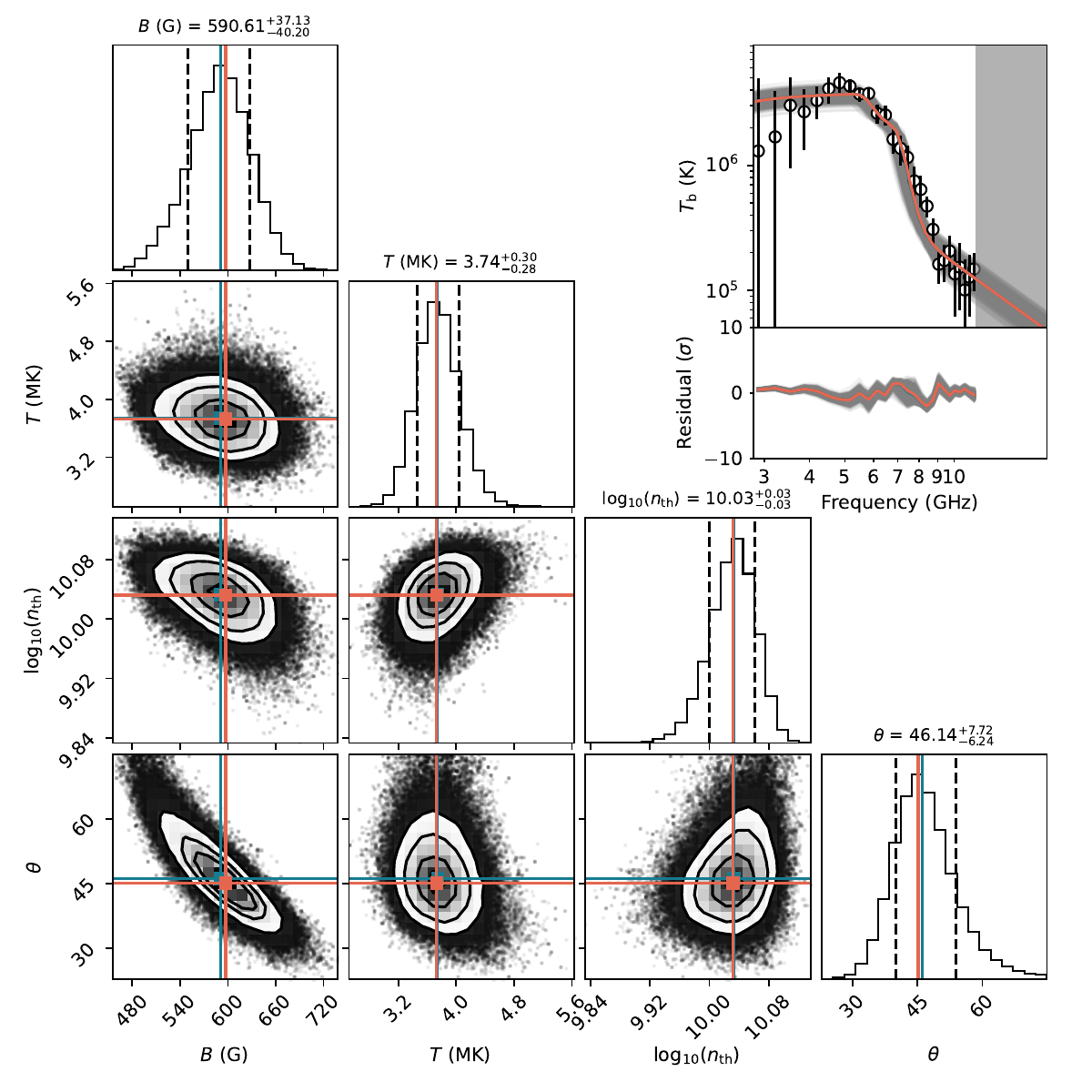}{1\textwidth}{}}
\caption{PPDs of the MCMC analysis for the spectrum from Region B-1 (17:26:36 UT). The red lines represent the fitting results by non-linear least chi-square method. While the blue lines are results from the MCMC simulation. The errors are the 1-{$\sigma$} confidence intervals of the PPDs. The gray curves in the upper right panels are spectra and residuals associated with results from MCMC samplings shown in the corner plot.}
\end{figure*}

\begin{figure*}[ht]
\gridline{\fig{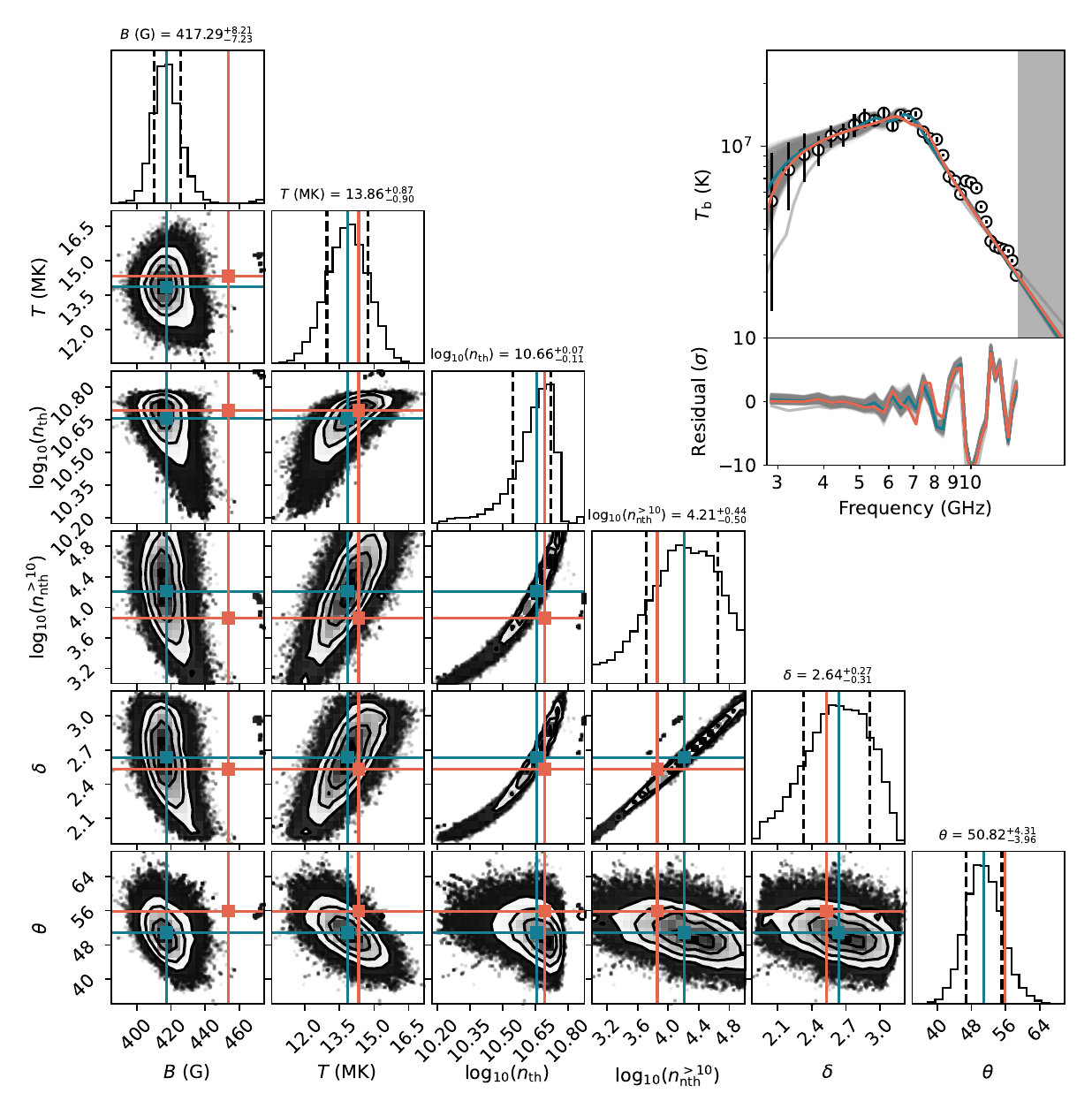}{1\textwidth}{}}
\caption{PPDs of the MCMC analysis for the spectrum from Region A-2 (17:29:46 UT). The red lines represent the fitting results by non-linear least chi-square method. While the blue lines are results from the MCMC simulation. The errors are the 1-{$\sigma$} confidence intervals of the PPDs. The gray curves in the upper right panels are spectra and residuals associated with results from MCMC samplings shown in the corner plot.}
\end{figure*}

\begin{figure*}[ht]
\gridline{\fig{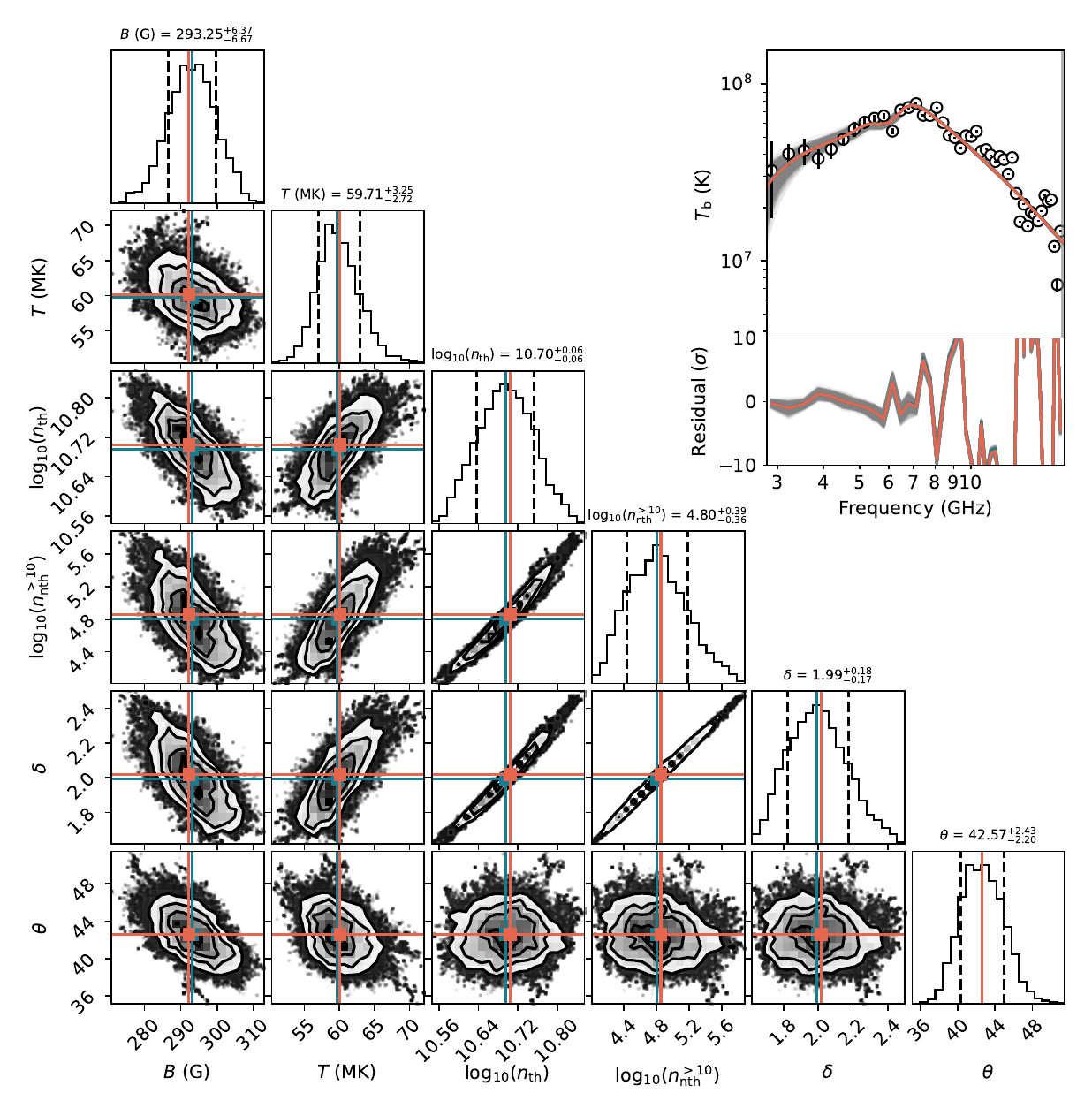}{1\textwidth}{}}
\caption{PPDs of the MCMC analysis for the spectrum from Region A-3 (17:31:23 UT). The red lines represent the fitting results by non-linear least chi-square method. While the blue lines are results from the MCMC simulation. The errors are the 1-{$\sigma$} confidence intervals of the PPDs. The gray curves in the upper right panels are spectra and residuals associated with results from MCMC samplings shown in the corner plot.}
\end{figure*}

\end{document}